# Proper Conformal Vector fields in Bianchi Type I Space-Times

Ghulam Shabbir

Faculty of Engineering Sciences

GIK Institute of Engineering Sciences and Technology

Topi Swabi, NWFP, Pakistan

Email: shabbir@giki.edu.pk

and

Shaukat Iqbal

Faculty of Computer Science and Engineering

GIK Institute of Engineering Sciences and Technology

Topi Swabi, NWFP, Pakistan

**Abstract**

Direct integration technique is used to study the proper conformal vector fields in non conformally flat Bianchi type-1 space-times. Using the above mentioned technique we have shown that a very special class of the above space-time admits proper conformal vector fields.

## 1. INTRODUCTION

This paper investigates the existance of proper conformal vector fields in Bianchi type-1 space-times by using the direct integration technique. The conformal vector field which preserves the metric structure upto a conformal factor carries significant interest in Einstein's theory of general relativity. It is therefore important to study these symmetries.

Throughout $M$ represents a four dimensional, connected, hausdorff space-time manifold with Lorentz metric $g$ of signature (-, +, +, +). The curvature tensor associated with $g_{ab}$, through the Levi-Civita connection, is denoted in component form by $R^a{}_{bcd}$, and the Ricci tensor components are $R_{ab} = R^c{}_{acb}$. The usual covariant, partial and Lie derivatives are denoted by a semicolon, a comma and the symbol $L$, respectively. Round and square brackets denote the usual symmetrization and skew-symmetrization, respectively.

Any vector field $X$ on $M$ can be decomposed as

$$X_{a;b} = \frac{1}{2}h_{ab} + F_{ab} \tag{1}$$

where $h_{ab}(=h_{ba}) = L_X g_{ab}$ and $F_{ab}(=-F_{ba})$ are symmetric and skew symmetric tensors on $M$, respectively. Such a vector field $X$ is called conformal vector field if the local diffeomorphisms $\psi_t$ (for appropriate $t$) associated with $X$ preserve the metric structure up to a conformal factor i.e. $\psi_t^* g = \phi g$, where $\phi$ is a nowhere zero positive function on $M$ and $\psi_t^*$ is a pullback map on $M$ [3]. This is equivalent to the condition that

$$h_{ab} = 2\phi g_{ab},$$

equivalently

$$g_{ab,c}X^c + g_{cb}X^c_{,a} + g_{ac}X^c_{,b} = 2\phi g_{ab}, \tag{2}$$

where $\phi : M \to R$ is the smooth conformal function on $M$, then $X$ is a called conformal vector field. If $\phi$ is constant on $M$, $X$ is homothetic (proper homothetic if $\phi \neq 0$) while $\phi = 0$ it is Killing [3]. If the vector field $X$ is not homothetic then it is called proper conformal. It follows from [3] that for a conformal vector field $X$, the bivector $F$ and the function $\phi$ satisfy (putting $\phi_a = \phi_{,a}$)

$$F_{ab;c} = R_{abcd}X^d - 2\phi_{[a}g_{b]c}, \tag{3}$$

$$\phi_{a;b} = -\frac{1}{2}L_{ab;c}X^c - \phi L_{ab} + R_{c(a}F_{b)}{}^c, \tag{4}$$

where $L_{ab} = R_{ab} - \frac{1}{6}R g_{ab}$.

## 2. Main Results

A Bianchi type-1 space-time is a spatially homogeneous space-time which admits an abelian Lie group of isometries $G_3$, acting on spacelike hypersurfaces, generated by the spacelike Killing vector fields which are [2,4]

$$\frac{\partial}{\partial x}, \frac{\partial}{\partial y}, \frac{\partial}{\partial z}. \tag{5}$$

The line element in usual coordinate system is [1]

$$ds^2 = -dt^2 + h(t)dx^2 + k(t)dy^2 + f(t)dz^2, \tag{6}$$

where $f, k$ and $h$ are some nowhere zero functions of $t$ only. The possible Segre type of the above space-time is $\{1,111\}$ or one of its degeneracies. A vector field $X$ is said to be a conformal vector field if it satisfy equation (2). One can write (2) explicitly using (6) we have

$$X^0_{,0} = \phi, \tag{7}$$

$$-X^0_{,1} + hX^1_{,0} = 0, \tag{8}$$

$$-X^0_{,2} + kX^2_{,0} = 0, \tag{9}$$

$$-X^0_{,3} + f X^3_{,0} = 0, \tag{10}$$

$$\dot{h} X^0 + 2h X^1_{,1} = 2h\phi, \tag{11}$$

$$hX^1_{,2} + kX^2_{,1} = 0, \tag{12}$$

$$hX^1_{,3} + f X^3_{,1} = 0, \tag{13}$$

$$\dot{k} X^0 + 2k X^2_{,2} = 2k\phi, \tag{14}$$

$$kX^2_{,3} + f X^3_{,2} = 0, \tag{15}$$

$$\dot{f} X^0 + 2f X^3_{,3} = 2f \phi. \tag{16}$$

Equations (7), (8), (9) and (10) give

$$\left.\begin{aligned}
X^0 &= \int \phi(t)dt + A^1(x,y,z) \\
X^1 &= A^1_x(x,y,z)\int \frac{dt}{h} + A^2(x,y,z) \\
X^2 &= A^1_y(x,y,z)\int \frac{dt}{k} + A^3(x,y,z) \\
X^3 &= A^1_z(x,y,z)\int \frac{dt}{f} + A^4(x,y,z)
\end{aligned}\right\}, \quad (17)$$

where $A^1(x,y,z), A^2(x,y,z), A^3(x,y,z)$ and $A^4(x,y,z)$ are functions of integration. In order to determine $A^1(x,y,z), A^2(x,y,z), A^3(x,y,z)$ and $A^4(x,y,z)$ we need to integrate the remaining six equations. To avoid details, here we will present only the result when the above space-time (6) admits proper conformal vector field. It follows from the above calculations; there exist only one possibility when the above space-time (6) admits proper conformal vector field which is:

**Case (1)** Four conformal vector fields:

In this case the space-time (6) becomes

$$ds^2 = -dt^2 + V^2(t)(e^{-2d_1 N(t)}dx^2 + e^{-2d_{11} N(t)}dy^2 + e^{-2d_{13} N(t)}dz^2) \quad (18)$$

and conformal vector field is

$$X^0 = V(t), \quad X^1 = d_1 x + d_2, \quad X^2 = d_{11} y + d_{12}, \\ X^3 = d_{13} z + d_{14}, \quad (19)$$

where $V(t) = \int \phi(t)dt + d_8$, $N(t) = \int \frac{1}{V(t)}dt$, $d_1, d_2, d_8, d_{11}, d_{12}, d_{13}, d_{14} \in R($ $d_1 \neq d_{11}, d_1 \neq d_{13}, d_{13} \neq d_{11}, d_1 \neq 0, d_{11} \neq 0, d_{13} \neq 0)$ and $\phi$ is no where zero function of $t$ only. The above space-time (18) admits four independent conformal vector fields in which three are Killing vector fields which are given in (5) and one is proper conformal vector field which is

$$Z = (V(t), d_1 x, d_{11} y, d_{13} z). \quad (20)$$

One can easily check that the above vector field (20) is not a homothetic vector field by substituting it into the homothetic equations.

Now consider the case when $d_{11} = d_{13}$, $d_{11} \neq d_1$ and the above space-time (18) becomes

$$ds^2 = -dt^2 + V^2(t)(e^{-2d_1 N(t)}dx^2 + e^{-2d_{11} N(t)}(dy^2 + dz^2)). \quad (21)$$

The above space-time admits five independent conformal vector fields in which four independent Killing vector fields which are: $\frac{\partial}{\partial x}, \frac{\partial}{\partial y}, \frac{\partial}{\partial z}$ and $z\frac{\partial}{\partial y} - y\frac{\partial}{\partial z}$ and one proper conformal vector field which is given in (20). The cases when $d_{11} = d_1$, $d_{11} \neq d_{13}$ and $d_1 = d_{13}$, $d_{11} \neq d_1$ are exactly the same.